# Absence of temperature-dependent phonon anomalies in $Sr_2IrO_4$ and $Sr_3Ir_2O_7$

K. Sen[1,*], R. Heid[1], S. M. Souliou[1], D. Boll[1], A. Bosak[2], N. H. Sung[3], J. Bertinshaw[3,4], H. Gretarsson[4], B. J. Kim[5,6], F. Weber[1], and M. Le Tacon[1,†]

[1]*Institute for Quantum Materials and Technologies, Karlsruhe Institute of Technology, 76021 Karlsruhe, Germany*
[2]*European Synchrotron Radiation Facility, BP 220, F-38043 Grenoble Cedex, France*
[3]*Max Planck Institute for Solid State Research, Heisenbergstrasse 1, D-70569 Stuttgart, Germany*
[4]*Deutsches Elektronen-Synchrotron DESY, Notkestraße 85, D-22607 Hamburg, Germany*
[5]*Department of Physics, Pohang University of Science and Technology, Pohang, South Korea*
[6]*Center for Artificial Low Dimensional Electronic Systems, Institute for Basic Science (IBS), Pohang, South Korea*

Following previous works reporting an anomalous behavior of several zone-center optical phonons across the magnetic transition of square lattice iridates $Sr_2IrO_4$ and $Sr_3Ir_2O_7$, we have investigated the lattice dynamics as a function of momentum in these materials by means of high-resolution inelastic x-ray scattering (IXS). The observed phonon energies and scattering intensities across the Brillouin zone are in excellent agreement with *ab*-initio lattice dynamical calculations based on non-magnetic density-functional-perturbation theory (DFPT). Our results do not evidence any renormalization of the phonons at finite momentum across the magnetic transition of $Sr_2IrO_4$. The only anomalous behavior was detected for the in-plane polarized longitudinal-acoustic phonon branch in $Sr_3Ir_2O_7$, which anomalously softens towards low temperatures and might be related to anisotropic negative thermal expansion in this compound. No anomalies related to potential charge ordering were observed.

## I. INTRODUCTION

The strong entanglement of the spin and orbital motions of the 5d electrons in iridium oxides has opened fresh perspectives in the quest for novel exotic electronic states in quantum materials. The effects of spin-orbit coupling (SOC) are particularly prominent in the Ir-based Ruddlesden-Popper series, $Sr_{n+1}Ir_nO_{3n+1}$ (n=1, 2 and ∞), which exhibit singular electronic and magnetic properties related to the formation of $J_{eff}$ = 1/2 pseudospins. For instance, despite moderate electronic correlations (rooted in the large spatial extension of the 5d electronic wave functions), a spin-orbit Mott insulating state forms in single-layer (n=1) $Sr_2IrO_4$ (Sr-214) as these pseudospins order antiferromagnetically [1,2]. A smaller Mott gap is observed in the bilayer $Sr_3Ir_2O_7$ (Sr-327) compound [3,4], whereas $SrIrO_3$ (n=∞) is a correlated semimetal exhibiting strange metal charge dynamics [5,6]. In many aspects, the phenomenology encountered in these layered iridates is reminiscent to that of high-temperature superconducting cuprates and sparked an intense scientific activity, further fueled by the recent observations of Fermi arcs [7], d-wave gap [8] or persistent high energy magnetic excitations upon electron-doping of the material [9].

As a corollary to the large orbital contribution to the pseudospin wave functions, the details of the pseudospin-pseudospin interaction, which in turn govern the low-energy physics of these systems, is highly dependent on the lattice geometry [10]. The electronic ground state of the iridates is in particular expected to sensitively depend on the octahedra rotation, which also provides attractive routes to electronic structure engineering using strain tuning via, e.g., epitaxial growth. Conversely, the crystal lattice is expected to respond to changes in the electronic or magnetic structure. This has been illustrated by the large anisotropic expansion of the unit cell induced by changes in the canting angle of Ir-moments of Sr-214 driven by electrical currents [11]. This has further reaching consequences as it was for instance recently demonstrated that the coupling of the pseudospins to the lattice is essential to account for the experimentally observed magnetic structure of these compounds [12].

Evidence for such pseudospin-lattice coupling has also been reported at the dynamical level in Sr-214 and Sr-327. Previous Raman scattering studies have revealed sizeable renormalization of several phonons across the magnetic ordering transition temperature [13–16]. Recent ultrafast optical spectroscopic studies of coherent phonons in Sr-327 confirmed unusually large pseudospin-lattice coupling in this material [17]. All these optical studies are however limited to the impact of magnetism on the zone-center phonon modes. In order to obtain a full overview of the momentum dependence of the pseudospin-phonon coupling in square lattice iridates, an investigation across the entire Brillouin zone is needed. Last but not least, charge density modulations breaking the four-fold into two-fold crystal symmetry have recently been inferred in electron-doped metallic Sr-327 [16,18]. Such modulations which have proven rather ubiquitous in condensed matter systems are very often, albeit not systematically [19], associated with temperature-dependent phonon anomalies [20–22] at finite

---

[*] Present address: Institute of Functional Interfaces, Karlsruhe Institute of Technology, 76021 Karlsruhe, Germany
[†] matthieu.letacon@kit.edu

momentum, which further motivate a systematic investigation of the lattice dynamics.

In this paper, we report on a momentum dependent lattice dynamics study of Sr-214 and Sr-327. We have used high resolution inelastic x-ray scattering (IXS) to map the dispersion of selected phonons above and below the magnetic ordering transition temperature $T_N$. Our primary goal was to focus on the branches of the optical phonons that displayed the largest renormalization at the zone center [14,15]. To this aim, we have performed *ab initio* lattice dynamical calculations based on non-magnetic density-functional-perturbation theory (DFPT), which enabled us to select the Brillouin zones that maximized the scattering intensity of these modes against the others. Excellent agreement is found between our experimental observations and the intensities calculated for a non-magnetic ground state with optimized lattice parameters. Within experimental accuracy, we did not observe any renormalization of the investigated optical phonons at finite momentum across the AF transition in Sr-214, nor of the acoustical branches in this material.

In the explored regions of the reciprocal space, no dispersion anomalies in relation to possible charge ordering could be evidenced. We nevertheless detected an anomalous softening of the entire longitudinal-acoustic phonon branch of Sr-327 with decreasing temperature, which could relate to a recent report of anisotropic negative thermal expansion for this material [17].

## II. EXPERIMENTAL DETAILS

### A. Crystal growth and lattice parameters

Single crystals of Sr-214 and Sr-327 were grown with flux methods as described in ref. [23]. The typical size of the crystals was ~ $0.2 \times 0.15 \times 0.05$ mm$^3$. Crystals were characterized by energy-dispersive x-ray spectroscopy, single-crystal x-ray diffraction, resistivity and magnetization measurements [23].

### B. Inelastic x-ray scattering

The inelastic x-ray scattering (IXS) experiments were carried out at the ID28 beamline of the European Synchrotron Radiation Facility (ESRF). The incident photon energy was 17.794 keV and the corresponding instrumental resolution was ~2.8 meV (full width at half maximum determined from the fitting of a resolution limited spectra by a pseudo-voigt lineshape). The x-ray beam was focused by multilayer mirrors to a spot of ~$50 \times 30$ μm$^2$ on the sample surface. The momentum transfer was selected by the crystal orientation and the scattering angle. The momentum resolution was set to ~0.025 Å$^{-1}$ in the horizontal scattering plane and 0.075 Å$^{-1}$ perpendicular to it. Energy scans were collected at constant momentum transfer $\boldsymbol{Q} = \boldsymbol{\Gamma_{hkl}} + \boldsymbol{q}$, where $\boldsymbol{\Gamma_{hkl}}$ is a reciprocal lattice point and $\boldsymbol{q}$ is the reduced wave vector. The components $(Q_h, Q_k, Q_l)$ of the scattering vectors are expressed in reciprocal lattice units (r.l.u.) $(Q_h, Q_k, Q_l) = (h\frac{2\pi}{a}, k\frac{2\pi}{b}, l\frac{2\pi}{c})$ with the lattice settings of $a = b = 5.48$ Å, c = 25.80 Å (space group $I4_1/acd$) [24] and a = 5.522 Å, b= 5.521 Å, c = 20.917 Å (space group $Bbca$) [3] for Sr-214 and Sr-327, respectively, at room temperature. The crystal temperature was adjusted between 100 and 350 K using an Oxford Cryostream 700 Plus. The phonon peaks in the collected IXS spectra were fitted using standard damped harmonic oscillator (DHO) functions convoluted with the experimental resolution.

## III. COMPUTATIONAL DETAILS

*Ab initio* lattice dynamical calculations were performed based on DFPT using the mixed-basis pseudopotential method. We employed the same type of pseudopotentials, local functions and convergence parameters as in our previous work on SrIrO$_3$ [6]. The exchange correlation functional is represented by the local-density approximation (LDA) [25]. No magnetic order was considered, but spin-orbit coupling was included consistently in all calculations. The crystal structures were optimized within LDA prior to lattice dynamical calculations. Phonons of Sr-214 were obtained for the distorted perovskite structure corresponding to a $\sqrt{2} \times \sqrt{2} \times 2$ supercell (tetragonal space group $I4_1/acd$) [24]. The optimized lattice parameters are a= b= 5.3678 and c= 25.499 Å. In the case of Sr-327, phonon calculations were performed for the slightly-distorted orthorhombic structure with space group C2/c [26]. Here, the structural optimization resulted in lattice parameters a= 20.603, b= 5.4144 and c= 5.3895 Å. Brillouin zone integrations were performed using *k*-point meshes of $12 \times 12 \times 2$ and $2 \times 8 \times 8$ for Sr-214 and Sr-327, respectively, in conjunction with a Gaussian smearing of 0.2 eV. We would like to point out that theoretical optimization underestimated the experimental crystal unit cell volumes by 5.2 and 5.1 % for Sr-214 and Sr-327, respectively. This is in line with the frequently observed tendency of LDA to overbind, typically leading to a unit cell volume of 3 to 6 % smaller than the experimental one. In contrast, the calculated phonon energies are in good agreement with the experiment [27], whereas the use of the experimental lattice parameters results in an underestimation of the phonon energies of about 12%. DFPT calculations of dynamical matrices were performed on tetragonal and orthorhombic $2 \times 2 \times 2$ momentum meshes for Sr-214 and Sr-327, respectively. Standard Fourier interpolation of the dynamical matrices then provided the eigenvectors and eigenfrequencies of vibrational modes at arbitrary momenta, which enter the structure factor calculations. For comparison with the measured intensities, the structure factors were combined with DHO functions convoluted with experimental resolution.

## IV. RESULTS

### A. Sr$_2$IrO$_4$

In Sr-214, the magnetic moment of the Ir-atoms lies on the basal plane with magnetic moments of 0.20 and 0.05 μ$_B$/Ir-atom along the *a*- and *b*-axis, respectively [28]. Earlier, a Raman scattering study demonstrated that a zone-center mode with the A$_{1g}$ symmetry at ~ 23 meV displays a significant asymmetry in the paramagnetic state ($T > T_N$=240 K), which was attributed to the coupling between the lattice and the underlying pseudospin fluctuations. This asymmetry completely disappears in the AF state below $T_N$ [14,15]. Our structure factor calculation indicates that the dispersion of this particular mode along the (1, 1, 0) reciprocal direction, which corresponds to the in-plane direction of the magnetic ordering wavevector in the considered unit cell, shows a sizeable intensity in the Brillouin zone (BZ) centered on the $\Gamma_{0016}$ Bragg reflection, which also selects a *c*-axis polarized transverse optical phonon. We have also investigated the longitudinal acoustical phonon (as well as a couple of in-plane polarized vibrations) starting from the $\Gamma_{220}$ BZ center.

FIG. 1 (a) shows the imaginary part of dynamic susceptibility $\chi''(\mathbf{Q},\omega)$ of Sr-214 as a function of energy transfer ($\omega$) and momenta along the (1, 1, 0) starting from $\Gamma_{220} = (2, 2, 0)$ at $T = 200$ K. $\chi''(\mathbf{Q},\omega)$ is obtained by dividing the IXS intensity with the temperature dependent Bose-factors after subtracting the elastic line from the IXS spectra. As seen by comparison with FIG. 1 (b) in which the calculated scattering intensity is reported, the overall experimental momentum dependence of the scattered intensity is very well accounted for by our non-magnetic DFPT calculation. In particular, the number of phonon modes at all $\mathbf{Q}$, as well as their energies and relative intensities are well reproduced by the calculation. This is further best illustrated by plotting representative $\chi''(\omega)$ spectra at a few $\mathbf{Q}$ at $T$ = 200 and 295 K, as shown in FIG. 1(c) along with the calculated $\chi''(\omega)$.

The experimental and computational results for the c-axis polarized modes propagating along (1, 0, 0) in the BZ centered on $\Gamma_{0016}$ are shown FIGS. 1 (d)-(f)). Again, the phonon modes and their relative intensities are well reproduced by our calculations. In addition to an intense transverse acoustical (TA) phonon, we do observe two optical modes above 20 meV, one of which corresponds to the aforementioned Raman-active $A_{1g}$ mode and displays the highest intensity, close to 24.5 meV.

As can be seen from FIG. 2 in which we summarize the energy of the phonon modes across the BZ as determined from the fits at 200 (below $T_N$) and 295K (above $T_N$) together with the calculated dispersion of the relevant modes, we did not observe a sizeable renormalization of any of the measured phonon energies across the magnetic transition. Furthermore, within the resolution of our experiment, no change was found in their linewidths across $T_N$. More specifically, no broadening of the $A_{1g}$ phonon could be resolved at finite momentum above $T_N$. Finally, the relative intensities of the phonon modes remain the same across $T_N$, as shown in FIGS. 1(c) and (f), indicating that the magnetism in Sr-214 has a negligible impact on the oscillator strengths of these phonons at finite wave vectors $\mathbf{q}$. This confirms and extends the conclusion from a previous report of an independent IXS study for the in-plane modes in Sr-214 around $\Gamma_{450}= (4, 5, 0)$ [29].

### B. Sr$_3$Ir$_2$O$_7$

We have used a similar approach to investigate the lattice dynamics in the bilayer compound Sr-327, in which we have measured acoustic phonons near the Brillouin zone center $\Gamma_{400} = (4, 0, 0)$. Longitudinal acoustic (LA) phonons at 100 K < $T_N$ are shown in FIG. 3(a), and transverse c-axis polarized acoustical and optical modes measured along the (-x,0,17) reciprocal direction in FIG. 3(c).

Again, the latter choice of Brillouin zone was motivated by the fact that the DFPT structure factor calculations predicted sizeable structure factors of the Raman-active (zone-center) $A_{1g}$ mode at ~ 23 meV which exhibits an asymmetric lineshape in the paramagnetic state [14]. The calculated phonon intensity is shown in FIGS. 3(b) and 3(d), respectively, and as in the case of Sr-214, reproduces well the relative intensities of the observed modes. Note however that in the case of c-axis polarized modes, the low intensity recorded below ~15 meV does not enable reliable analysis of the TA branch (which exhibits an instability in the calculation, which does not affect our conclusions).

In contrast to Sr-214, we now observe a singular temperature dependence of the phonon energies, especially for the acoustical modes recorded next to $\Gamma_{400}$. This is best seen in FIGS. 4(a) and (b) in which we display representative $\chi''(\mathbf{Q},\omega)$ showing the LA phonon at $\mathbf{q}$ = (-0.2, 0, 0) and the TA mode at $\mathbf{q}$ = (0, 0.2, 0) respectively above (at 350 K) and below (100K) the magnetic ordering temperatures. We have extracted the phonon mode energies at these two temperatures by fitting the phonon modes with resolution-convoluted DHO profiles (solid lines in FIGS. 4(a) and (b)). The resulting phonon energies are further shown as a function of momentum transfer FIG. 4(c) along with the calculated dispersion lines. We observed a softening of the LA mode by ~0.3 meV when the sample is cooled down from 350 K to 100 K, whereas the TA mode hardens by ~0.3 meV, at least up to the momentum transfer of $\mathbf{q} \leq 0.3$. This can be seen in FIG. 4(d), where we show the relative energy difference of the two modes at these two temperatures ($\Delta E/E = (E_{100K} - E_{350K})/E_{350K}$). The temperature dependence of phonon energies generally relates to anharmonicity which also governs the thermal expansion of the lattice. As such, phonons are expected to harden with decreasing temperature as the lattice contracts upon cooling. For this reason, the softening of the LA mode towards low temperatures is surprising. Before discussing this in the next section, we turn to the behavior of the optical modes.

FIG. 4(e) shows the energy of the optical modes as a function of momentum transfers at 295 and 100 K, as obtained from resolution-convoluted DHO fits. For the most of the optical phonons, we observe a phonon hardening as expected from a regular lattice contraction towards low temperatures (see FIG. 4(f)). For instance, both optical phonon branches at $\mathbf{q}$ = (−0.5, 0, 1) yield A phonon hardening of ~2.7 %, when cooled down to 100 K from 295 K, consistent with estimates based on the volume contraction [see Appendix A] and with previous optical Raman scattering data on these samples [14].

### V. DISCUSSION

We start pointing out that in contrast to the behavior reported at the zone center in previous Raman scattering [13–16] or ultrafast optical [17] studies on both materials, no particular effect of the magnetic ordering could be resolved on the investigated optical phonons. Each time the linewidths appear resolution limited, and no lineshape renormalization across the transition could be observed. The fact that the intensity and dispersion of these modes are well described by a non-magnetic calculation both above and below the magnetic transition temperatures further indicates that the pseudospin to phonon coupling effects are strongly momentum dependent and essentially confined to the low momentum transfers.

The most surprising result of this study is certainly the anomalous behavior of the in-plane polarized acoustic phonons in Sr-327. In particular, the LA branch shows a small but measurable softening between 350 and 100 K, while the TA branch propagating along (1, 0, 0) hardens. We note that the strength of these effect appears stronger at low momentum transfer. As pointed out earlier, from regular lattice anharmonicity, hardening of phonons is expected upon cooling as the unit cell shrinks. To date, there are only very few crystallographic studies of Sr-327 [3,26]. Only a subset of those have investigated the low temperature crystal structure, and the results appear at times contradictory. Using neutron diffraction, Hogan et al. report a contraction of all lattice parameters between room temperature to 100K [26]. Earlier powder X-ray diffraction from the same group [4] however reported an anisotropic thermal contraction of this material, which expands along the c-axis upon cooling but contracts in the basal plane. Along these lines, in a more recent ultrafast optical study which confirmed the strong coupling of the $A_{1g}$ phonons to the pseudospins [17], the evolution of the *c*-axis lattice parameter as function of temperature was investigated using x-ray diffraction and revealed its anomalous expansion upon cooling across the magnetic transition. We note that there are significant discrepancies in the reported absolute value of the *c*-axis lattice parameter which calls for a more systematic structural investigation.

In this context, we nevertheless note that our observation of the softening of the LA branch upon cooling supports, at least qualitatively, the possibility of an expansion of some of the lattice parameters upon cooling. The thermal expansion coefficient is indeed directly proportional to the average Grüneisen parameter $\gamma$, obtained after weighting each individual phonon Grüneisen parameter ($\gamma_i = -E_i/\Delta E_i \cdot \Delta V/V$) by its contribution to the heat capacity.

Based on ref. [26], we can estimate the contraction of the Sr-327 unit cell between 350 and 100K to be of the order of -0.6 % ($\Delta V/V = (V_{100K} - V_{350K})/V_{350K}$) [30]. The observed softening of the LA branch by 6 % in this temperature range (FIG. 4(a)), yields a negative Grüneisen parameter, $\gamma_{LA} \approx -10$. This is anomalous both in amplitude (typical Grüneisen parameters are of the order of 2-3 [31]) and in sign. The hardening of the TA mode also gives rise to a fairly large $\gamma_{TA} \approx 9$. This indicates that the temperature-dependence of both LA and TA modes in Sr-327 cannot be attributed to usual lattice anharmonicity effects, suggesting a possible connection to the anomalous contraction of the *c*-axis lattice parameter upon cooling reported in ref. [17].

Finally, we comment on the recently suggested appearance of four-fold symmetry breaking charge density modulations in electron-doped metallic Sr-327 [16,18]. Neither in Sr-214 nor in Sr-327 have we evidenced a localized softening or broadening of any of the phonon lines at finite momentum, at least in the explored regions of the reciprocal space.

## VI. CONCLUSION

We studied momentum-resolved phonons in antiferromagnetic (AF) single crystals of single-layer Sr$_2$IrO$_4$ (Sr-214) and bilayer Sr$_3$Ir$_2$O$_7$ (Sr-327) with inelastic x-ray scattering (IXS) across their AF transitions. We have supplemented our experimental results with non-magnetic DFPT calculations for both compounds. Such non-magnetic DFPT calculations describe the experimental results reasonably well with respect to phonon energies and structure factors.

Within the experimental accuracy of the present experiment, carried out with energy resolution of ~3 meV, we could not identify any change in the linewidths of the phonons across $T_N$ in Sr-214 and Sr-327. Earlier, Raman scattering studies that reported the asymmetric broadening of zone-center $A_{1g}$ modes in both compounds due to the coupling of the lattice and pseudospin fluctuations were carried out with much better energy resolutions (typically 0.1 to 0.2 meV), which remain challenging to obtain in IXS experiments. We obtained anomalously large Grüneisen parameters ($\gamma$) with $|\gamma|\sim10$ for both the LA and the TA branches propagating along the [100] direction in Sr-327. Moreover, the LA branch shows a phonon softening towards low-temperatures, i.e. the corresponding $\gamma$ is negative, which could be related to negative thermal expansion in Sr-327. We did not observe any charge modulation related phonon anomaly in the investigated compounds, but given that those are notoriously dependent on the charge carrier concentration (in particular in the cuprates) additional investigation as a function of doping are desired.

## ACKNOWLEDGMENTS

R.H. acknowledges support by the state of Baden-Württemberg through bwHPC.

## APPENDIX A: Estimation of Grüneisen parameters for optical phonon modes in Sr-214 and Sr-327

An Earlier Raman scattering study of zone-center optical phonons shows that the modes of B$_{2g}$ symmetry does not show any sizeable coupling to magnetic degrees of freedom in Sr-214 and Sr-327. Rather, these modes follow a regular temperature-dependence arising from the conventional lattice anharmonic effects [14]. For this reason, we estimated the Grüneisen parameters of these modes using the temperature-dependent unit cell volume [26] and Raman scattering data, as summarized in Table I.

| Sample | T (K) | E (cm$^{-1}$) | V (Å$^3$) | $\gamma$ |
|---|---|---|---|---|
| Sr-214 | 300 | 385.5 | 779.00 | 8.9 |
|  | 200 | 390.0 | 777.98 |  |
| Sr-327 | 300 | 388.5 | 636.60 | 4.6 |
|  | 100 | 397.0 | 633.60 |  |

TAB. I. Grüneisen parameters ($\gamma$) calculated using temperature-dependent unit cell volume (V) and frequency (E) of Raman-active B$_{2g}$ phonon modes. The formula used for $\gamma$ is $\Delta E/E = -\gamma \, \Delta V/V$.

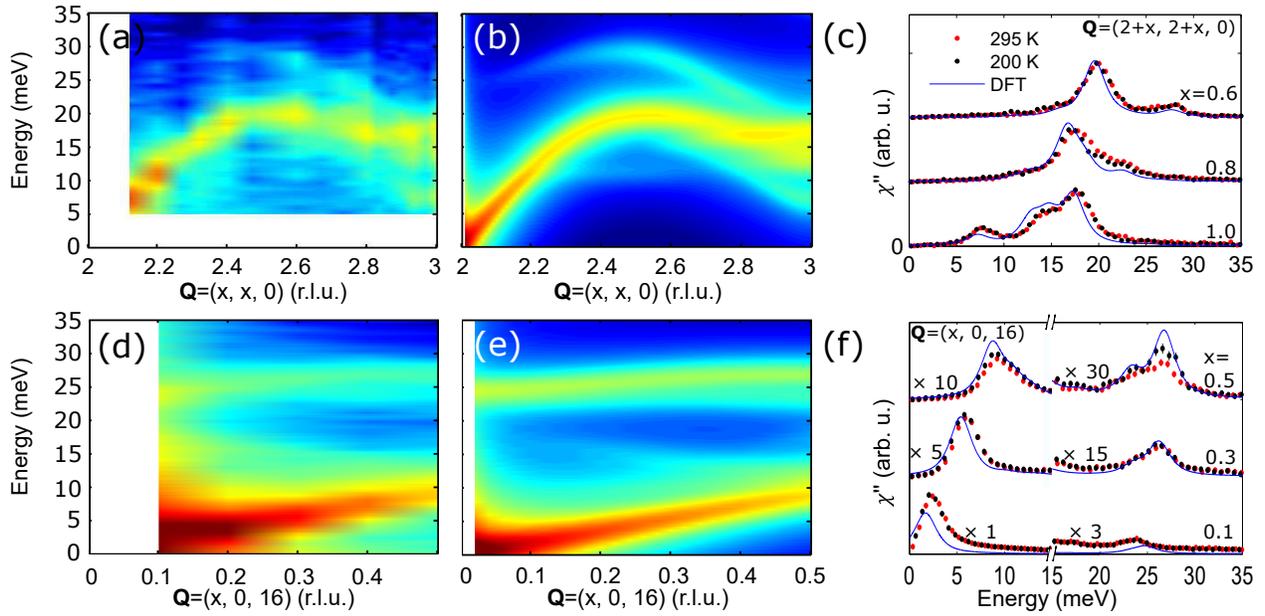

FIG. 1. (a) Experimental structure factors $\chi''(\mathbf{Q},\omega)$ of Sr$_2$IrO$_4$ next to the $\mathbf{\Gamma_{220}} = (2,2,0)$ zone center at 200 K. (b) Non-magnetic DFPT calculations of the corresponding $\chi''(\mathbf{Q},\omega)$. (c) Representative Bose-factor corrected inelastic x-ray scattering spectra (symbols) along $(1,1,0)$ from the $\mathbf{\Gamma_{220}}$ at 295 and 200 K for Sr$_2$IrO$_4$. The solid lines through the data are calculations based on DFPT. (d) Experimental $\chi''(\mathbf{Q},\omega)$ next to $\mathbf{\Gamma_{0016}}$ at 200 K. Note that to enable the visualization of both acoustical and optical phonons in panels (a)-(d) a logarithmic color scale is used. (e) The corresponding $\chi''(\mathbf{Q},\omega)$ obtained from non-magnetic DFPT calculations. (f) Representative IXS spectra (symbols) and the corresponding calculations (solid lines) for selected momenta along the $(1,0,0)$ direction from $\mathbf{\Gamma_{0016}}$.

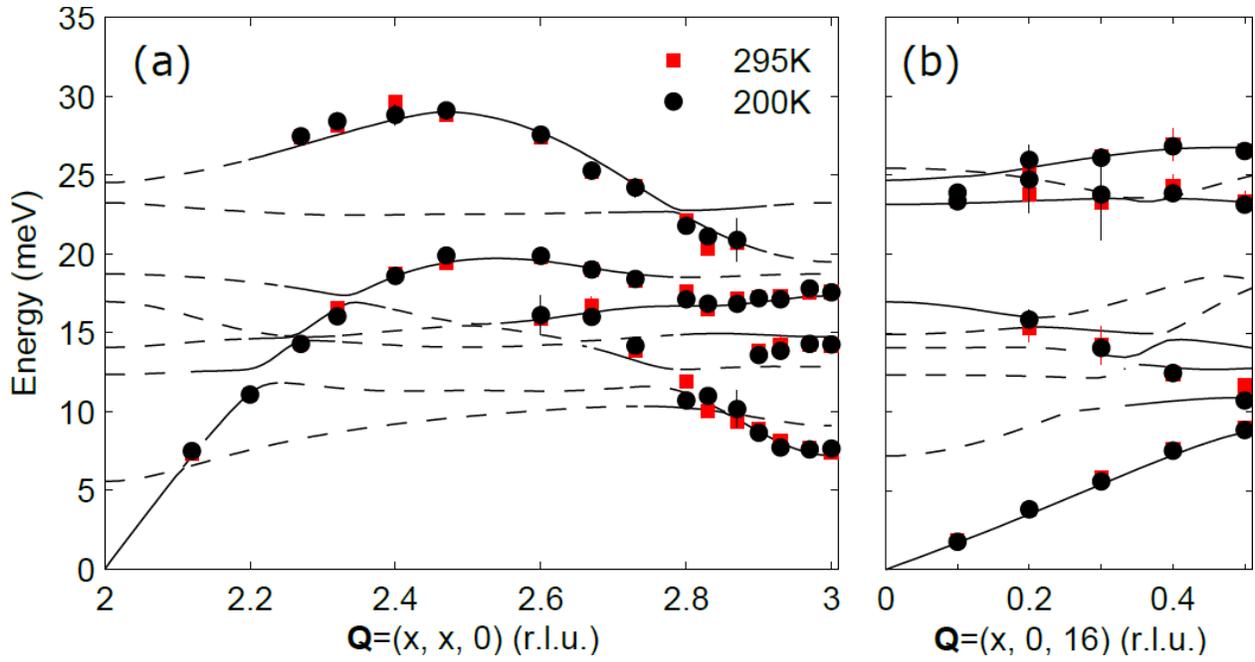

FIG. 2. (a) Phonon dispersion (symbols) along the (1, 1, 0) direction at T =295 and 200 K for $Sr_2IrO_4$ measured from the $\Gamma_{220}$ zone center. The relevant phonon dispersion branches from DPFT are also given. (b) Dispersion of the phonons measured from the $\Gamma_{0016}$ along the (1, 0, 0) direction. In both panels, the plain lines represent the modes with a sizeable calculated structure factor in the considered Brillouin zone (dashed lines correspond to vanishing structure factor).

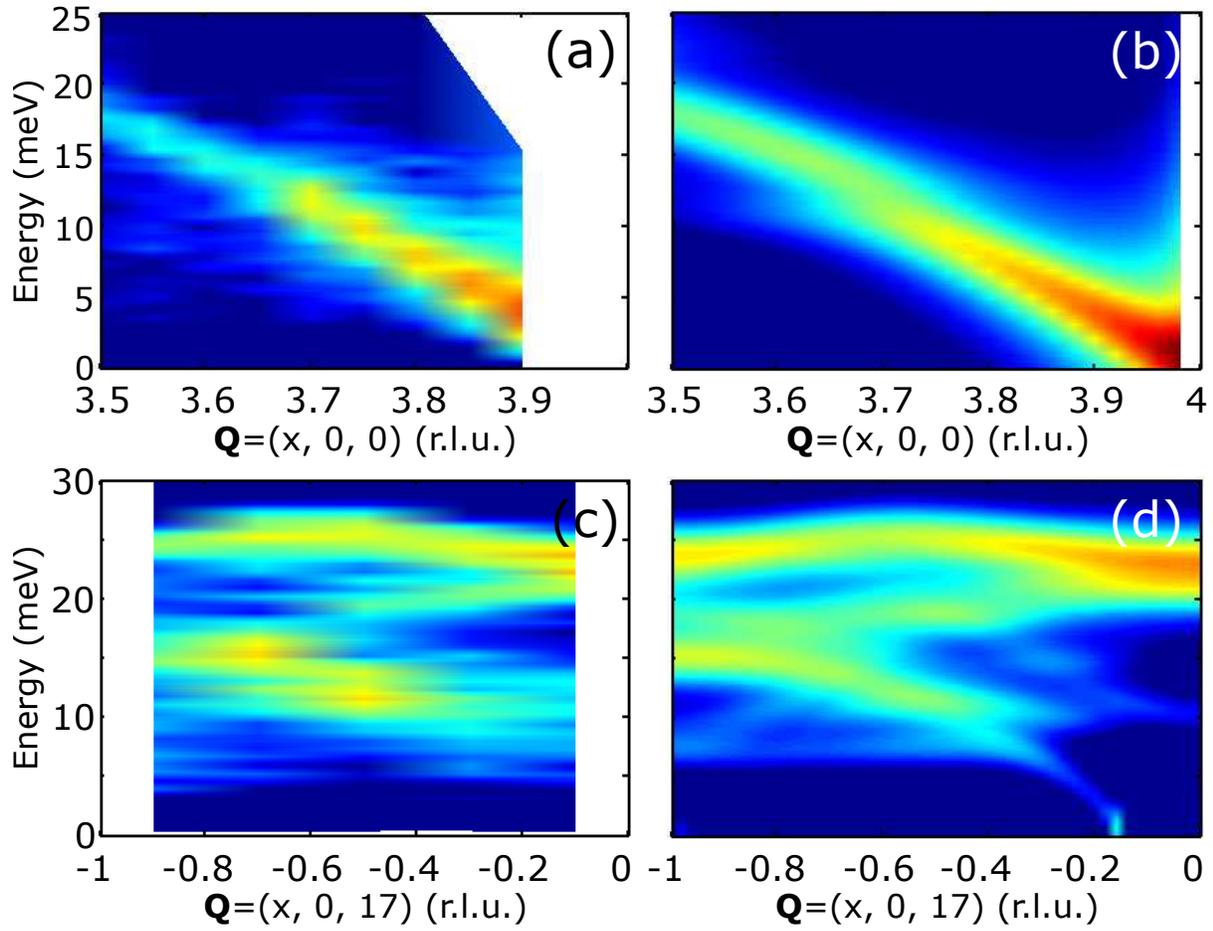

FIG. 3. (a) The experimental $\chi''(\mathbf{Q}, \omega)$ of the longitudinal acoustic branch of Sr$_3$Ir$_2$O$_7$ measured from $\Gamma_{400}$ at 100 K. (b) The corresponding calculation of $\chi''(\mathbf{Q}, \omega)$. (c) Experimental $\chi''(\mathbf{Q}, \omega)$ of the optical branches measured next to $\Gamma_{0017}$ at 100 K. (d) The corresponding calculation of $\chi''(\mathbf{Q}, \omega)$. Note that to enable the visualization of all modes in a single plot a logarithmic color scale is used.

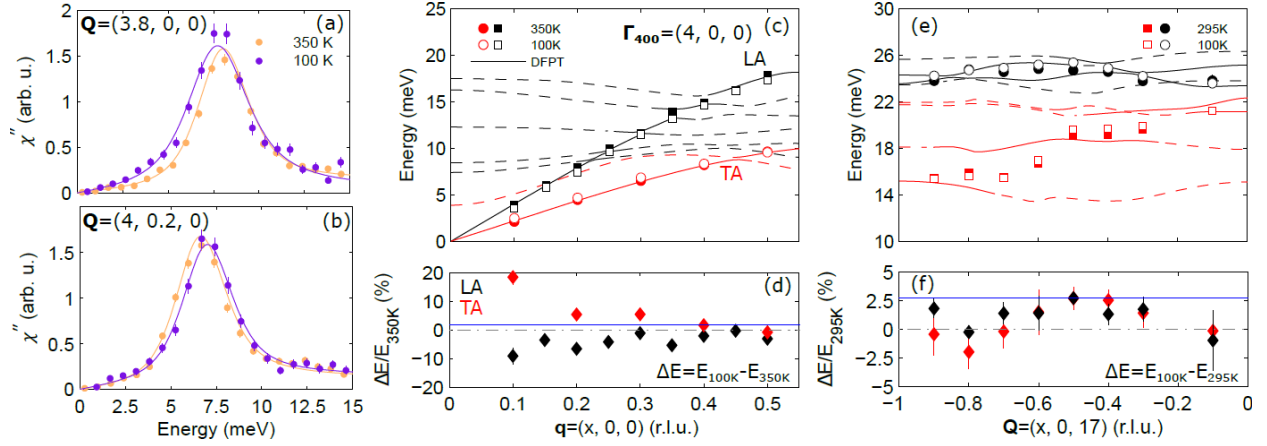

FIG. 4. Bose-factor corrected representative IXS spectra (symbols) and the corresponding resolution-convoluted damped harmonic oscillator fits (lines) for the acoustic phonon modes of $Sr_3Ir_2O_7$ measured next to the BZ center $\Gamma_{400}$. (a) Longitudinal acoustic (LA) mode measured at $\mathbf{Q} = \Gamma_{400} - (0.2, 0, 0)$. (b) Transverse acoustic (TA) mode measured at $\mathbf{Q} = \Gamma_{400} + (0, 0.2, 0)$. (c) Phonon dispersions for the LA and TA branches at 350 and 100 K. The relevant phonon branches from the DFPT calculations are assigned. The modes with sizeable structure factor are plotted with continuous line. (d) Relative energy difference of the phonon modes at these two temperatures. The solid horizontal blue line indicates the expected energy difference for a typical Grüneisen parameter of $\gamma = 3$ due to regular lattice anharmonicity. (e) Optical phonon dispersions near at 295 and 100 K. The relevant phonon branches from DFPT calculations are assigned. The modes with sizeable structure factor are plotted with continuous line. (f) Energy difference of the phonon modes at these two temperatures in percentage. The solid horizontal blue line indicates the expected energy difference for $\gamma = 4.6$, as obtained from Raman scattering experiments (see APPENDIX A).